\documentclass[twoside]{article}
\usepackage{amsmath}
\usepackage{amssymb}
\usepackage{cite}
\usepackage{graphicx}

\input ijmpblat
\newcommand{\nn}{\nonumber}

\newcommand{\etal}{\textit{et al.}}
\newcommand{\kb}{k_{_{\mathrm{B}}}}
\newcommand{\eps}{\varepsilon}
\newcommand{\bp}{\mathbf{p}}
\newcommand{\la}{\left<}
\newcommand{\ra}{\right>}
\newcommand{\up}{\uparrow}
\newcommand{\dn}{\downarrow}
\newcommand{\Tc}{\ensuremath{T_{\mathrm{c}}}}
\newcommand{\jour}[4]{{\nineit #1}\ {\ninebf #2},\ #3\ (#4)}
\newcommand{\Fref}[1]{Fig.~\ref{#1}}
\newcommand{\Eref}[1]{Eq.~(\ref{#1})}
\newcommand{\Rref}[1]{Ref.~\citen{#1}}
\newcommand{\PRL}{Phys. Rev. Lett.}
\newcommand{\PR}{Phys. Rev.}

\begin{document}
\runninghead{T. Mishonov \& E. Penev}%
            {Thermodynamics of anisotropic-gap and multiband
         clean BCS superconductors}
%
\thispagestyle{empty}
\setcounter{page}{1}
\copyrightheading{}
\vspace*{0.88truein}
\fpage{1}
\centerline{\textbf{THERMODYNAMICS OF ANISOTROPIC-GAP AND}}
\vspace*{0.035truein}
\centerline{\textbf{MULTIBAND CLEAN BCS SUPERCONDUCTORS}}
\vspace*{0.37truein}
\centerline{\footnotesize TODOR MISHONOV$^{\dag,\ddag,}$\footnote{
   Corresponding author; phone: (++32) 16 327183, fax: (++32) 16 327983,\\
   e-mail:~\texttt{todor.mishonov@fys.kuleuven.ac.be}
               }\ \
  {\normalsize and} EVGENI PENEV$^{\ddag}$
}
\vspace*{0.015truein}
\centerline{$^{\dag}$\footnotesize\it
   Laboratorium voor Vaste-Stoffysica en Magnetisme, %
   Katholieke Universiteit Leuven
}
\baselineskip=10pt
\centerline{\footnotesize\it
   Celestijnenlaan 200 D, B-3001 Leuven, Belgium
}
\vspace*{0.015truein}
\centerline{$^\ddag$\footnotesize\it
   Faculty of Physics, Sofia University ``St. Kliment Ohridski''
}
\baselineskip=10pt
\centerline{\footnotesize\it
    5~J.~Bourchier Blvd., 1164 Sofia, Bulgaria
}
\vspace*{0.225truein}

\vspace*{0.21truein}
\abstracts{The free energy, non-gradient terms of the Ginzburg-Landau
(GL) expansion, and the jump of the specific heat of a multiband
anisotropic-gap clean BCS superconductor are derived in the framework of a
separable-kernel approximation. Results for a two-band superconductor,
$d$-wave superconductor, and some recent models for MgB$_2$ are worked
out as special cases of the general approach. The classical results
for the GL coefficients are derived in a simple way, directly from the
general expression for the free energy of a BCS superconductor.}{}{}

\vspace*{0.21truein}
\keywords{Ginzburg-Landau theory, specific heat,
gap anisotropy, exotic superconductors}

\vspace*{1pt}\textlineskip
\section{Introduction}
\label{sec:1}
\vspace*{-0.5pt}
\noindent

The Landau theory of second-order phase transitions\cite{LLV} and its
realization for superconductors, the Ginzburg-Landau (GL) gauge
theory,\cite{GL} can be classified as belonging to the most
illuminating theoretical achievements in XXth-century physics. The
basic concepts advanced in these theories often find applications in
interdisciplinary research. The microscopic Bardeen-Cooper-Schrieffer
(BCS) theory\cite{bcs:57} makes it possible to calculate the
parameters of the GL theory. Thus the phenomenology of
superconductivity can be reliably derived once the parameters of the
microscopic Hamiltonian are specified. Such a scheme ensures that
there is no missing link between the microscopic theory and the
material properties of the superconductors.

The recent progress in physics of anisotropic-gap superconductors,
e.g., cuprates, borocarbides and MgB$_2,$ has attracted considerable
attention. However, the analysis of the thermodynamic behavior of
these compounds appeared to be erroneous in some cases, and often the
results from the classical papers in the field are not taken into
account. Thus, the present situation created the necessity that some
old theoretical results be delivered in a form suitable for fitting
the experimental data.

The purpose of the present paper is to provide a simple methodological
derivation of the non-gradient terms of the GL expansion. These are
then used to set up appropriate for numerical implementation
expressions which employ the electron band dispersion and momentum
dependence of the gap function at the Fermi surface. We have paid
particular attention to reflect the relation of our results to those
obtained in the pioneering papers in this field. Employing the
Bogoliubov-Valatin method\cite{Bogoliubov:58,Valatin:58} and the
standard expression for the free energy in the BCS model we present a
simple derivation of the results by Pokrovski\u\i\ and
Ryvkin\cite{Pokrovsky:63} and Gor'kov and
Melik-Barkhudarov.\cite{Gorkov:64} We have particularly focused on the
jump of the specific heat $\Delta C$ at the critical temperature \Tc\
which is expressed through the coefficients of the GL
expansion. Finally, the general formulae are applied to derive $\Delta
C$ in different models used to describe the superconducting cuprates,
borocarbides, and MgB$_2$.

\section{Microscopic formalism}
\label{sec:2}
\vspace*{-0.5pt}
\noindent

Our starting point is the \emph{total reduced}
Hamiltonian\cite{Moskalenko:59,Suhl:59,Ketterson:99} for a multiband
superconductor with singlet pairing
\begin{align}
 \hat{H}' &= \hat{H}_0 -\mu\hat{N} + \hat{H}_{\text{int}}
              \equiv \hat{H}'_0 + \hat{H}_{\text{int}} \nn\\
 & = \sum_{b,\bp,\alpha}\xi_{b,\bp}\,
     \hat{c}_{b,\bp\alpha}^{\dag} \hat{c}_{b,\bp\alpha}
   + \frac{1}{\cal{N}}\sum_{b,\bp}\sum_{b',\bp'} V_{b,\bp;b',\bp'}\,
     \hat{c}_{b,\bp\uparrow}^{\dag}\hat{c}_{b,-\bp\downarrow}^{\dag}
     \hat{c}_{b',-\bp'\downarrow}  \hat{c}_{b',\bp'\uparrow}.
\label{eq:H}
\end{align}
Here $\xi_{b,\bp}\equiv \eps_{b,\bp}-\mu$ is the quasi-particle
spectrum of the ``primed''\footnote{As usual it is convenient to work
at fixed chemical potential $\mu$ and introduce
$\hat{H}'_0=\hat{H}_0-\mu\hat{N},$ where $\hat{N}$ is the
quasiparticle number operator.}\ \ non-interacting Hamiltonian
$\hat{H}'_0,$ with $b$ being the band index, and
$\hat{c}_{b,\bp\alpha}^{\dag},$ $\hat{c}_{b,\bp\alpha}$ are the
quasi-particle creation and annihilation operators, respectively, for
the $b$th band, quasimomentum $\bp,$ and spin projection
$\alpha=\,\up,\dn.$ The second, four-fermion term in \Eref{eq:H} is
determined by the quasiparticle pairing interaction. The form of the
corresponding pairing matrix elements $V_{b,\bp;b',\bp'}$ has been one
of the vexing problems in the theory of high-\Tc\ superconductivity as
the underlying mechanism giving rise to pairing is still unknown.

However, once the interaction is specified, one can apply the standard
BCS treatment\cite{bcs:57,Ketterson:99} to obtain an equation for the
superconducting gap. For the Hamiltonian (\ref{eq:H}) this leads to
the BCS gap equation of the following familiar form
\begin{equation}
\Delta_{b,\bp} = \frac{1}{\cal{N}}\sum_{b',\bp'}(-V_{b,\bp;b',\bp'})
        \frac{1-2n_{b',\bp'}}{2E_{b',\bp'}}\Delta_{b',\bp'},
\label{eq:Gap}
\end{equation}
where
\begin{equation}
 E_{b,\bp} \equiv \sqrt{\xi_{b,\bp}^2 + |\Delta_{b,\bp}|^2}, \qquad
 n_{b,\bp}  =  \left[\exp(E_{b,\bp}/\kb T)+1\right]^{-1}
\end{equation}
are the quasiparticle energies, and the Fermi filling factors,
respectively, $\kb$ being the Boltzmann constant, and $T$ the
temperature. The summation over the band index $b'$ is restricted to
the partially filled (metallic) bands, containing the sheets which compose the Fermi
surface.

Generally, the BCS gap equation requires numerical treatment, but for
the special class of separable kernels $V$ it can be reduced to a
transcendental equation. This is the main reason why the gap symmetry
of some exotic high-\Tc\ superconductors is addressed on the basis of
model separable kernels. For specific pairing mechanisms, however, it
can be shown\cite{Mishonov:02} that the interaction kernel \emph{is}
of separable form. As a rule simple model estimates result in a sum of
such separable potentials. Therefore, as a next step we make the
\emph{separability ansatz}\cite{Markowitz:63} for the pairing
interaction kernel
\begin{equation}
  V_{b,\bp;b',\bp'} = - G \,\chi_{b,\bp}^*\,\chi_{b',\bp'},
\label{eq:kernel}
\end{equation}
where $G$ is a parameter, characteristic for the pairing interaction
process, and the momentum dependence is determined by the (generally
complex-valued) anisotropy function $\chi_{b,\bp}.$ Due to the
separable form of the BCS kernel (\ref{eq:kernel}) the temperature and
momentum dependences of the gap also factorize,
\begin{equation}
 \Delta_{b,\bp}(T)=\Xi(T)\,\chi_{b,\bp}.
\label{eq:anzatz}
\end{equation}
As emphasized by Gor'kov and Melik-Barkhudarov\cite{Gorkov:64} the
factorization of the superconducting gap into tem\-pe\-ra\-ture
($\Xi$) and momentum-dependent ($\chi$) multipliers is a general
result due to Pokrovski\u\i\ and Ryvkin,\cite{Pokrovsky:63} valid for
an arbitrary weak-coupling pairing interaction. The gap anisotropy
function is given by the eigenfunction of the pairing kernel
corresponding to the lowest eigenvalue. The factorable kernel
\Eref{eq:kernel} is just an interpolating Hamiltonian resulting in the
same thermodynamic behavior as that determined by the initial
kernel. Hence in what follows, without loss of generality, we shall
use a separable interaction. Such a type of interaction, however,
arises in lattice multiband models if one considers only local
(single-site) interactions.

Substituting \Eref{eq:anzatz} into \Eref{eq:Gap} and
passing from summation (for $\cal N\rightarrow \infty$) to integration
according to the general rule in the $D$-dimensional case
\begin{equation}
 \frac{1}{\cal N}
 \sum_{\bp} f(\bp) = \int_0^{2\pi}\!\!\dots\int_0^{2\pi}
 \frac{d\bp}{(2\pi)^D}\, f(\bp) \equiv \la f \ra_{\bp},
\label{eq:mean}
\end{equation}
we obtain a simple equation for the temperature dependence of the gap,
\begin{equation}
 G\sum_b\int_0^{2\pi}\!\!\dots\int_0^{2\pi}
 \frac{|\chi_{b,\bp}|^2}{2E_{b,\bp}}\,
 \tanh\!\left(\frac{E_{b,\bp}}{2\kb T}\right)\,
 \frac{d\bp}{(2\pi)^D} = 1,
\label{eq:sdGap}
\end{equation}
where the quasiparticle spectrum in expanded form reads
\begin{equation}
 E_{b,\bp} = \left[(\eps_{b,\bp} - E_{\mathrm{F}})^2 + |\Xi(T)|^2\,|\chi_{b,\bp}|^2
 \right]^{1/2},
\end{equation}
with $E_{\mathrm{F}}\equiv\mu$ being the Fermi energy. For the ease of
the following discussion we shall consider the simplest case of a
single band and suppress the band index. This does not entail any
restriction on the generality of the derivation.
\pagebreak[3]

\section{Thermodynamic properties}
\label{sec:GL}

In order to implement the GL idea for representing the free energy as
a function of the superconducting order parameter, $F(\Xi)$, we shall
employ the Bogoliubov-Valatin variational
approach.\cite{Bogoliubov:58,Valatin:58} Let us recall the main
framework of this standard procedure. As a first step one carries out
a transformation of the variables, introducing new Fermi operators
(for simplicity we will consider real kernels but the appropriate
generalization can be easily performed):
\begin{equation}
 \left(\begin{array}{c}
    \hat\psi^\dagger_{\bp\uparrow}\\
    \hat\psi_{-\bp\downarrow}
    \end{array}\right) \equiv
    \left(\begin{array}{cc}
        \cos\theta_{\bp}&\sin\theta_{\bp}\\
        -\sin\theta_{\bp}&\cos\theta_{\bp}
    \end{array}\right)
    \left(\begin{array}{c}
        \hat c^\dagger_{\bp\uparrow}\\
        \hat{c}_{-\bp\downarrow}
    \end{array}\right).
\end{equation}
Defining  the BCS ground state
\begin{equation}
 \left| \mathrm{BCS} \right. \rangle
 =\prod_{\bp}\left(\cos\theta_{\bp}+\sin\theta_{\bp}\,
 \hat c^\dagger_{\bp\uparrow}\hat c^\dagger_{-\bp\downarrow}
 \right)\left|0\right. \rangle,
\label{eq:bcs0}
\end{equation}
it is easily verified that
$
 \hat\psi_{\bp\alpha}\left| \mathrm{BCS} \right. \rangle=0=
 \hat c_{\bp\alpha}\left| 0 \right. \rangle,
$
and
$
 \langle \left.\mathrm{BCS}\right|\mathrm{BCS}\rangle= 1 =\langle 0|0\rangle.
$
The superconducting gap function is then used for a suitable
parameterization of the $\theta_{\bp}$ angle,
%
$  \tan 2\theta_{\bp} = \Delta_{\bp}/\xi_{\bp}.$
%
Denoting $u_{\bp}\equiv\cos\theta_{\bp},$
$v_{\bp}\equiv\sin\theta_{\bp},$
\begin{eqnarray}
 u_{\bp}^2
 =\frac{1}{2}\left(1+\frac{\xi_{\bp}}{E_{\bp}}\right),\qquad
 v_{\bp}^2 =\frac{1}{2}\left(1-\frac{\xi_{\bp}}{E_{\bp}}\right),
\end{eqnarray}
it is straightforward to verify that the condensation amplitude
$2u_{\bp}v_{\bp}$ and coherence factor $u_{\bp}^2-v_{\bp}^2$ read,
respectively,
\begin{equation}
 2u_{\bp}v_{\bp}=\sin
 2\theta_{\bp}=\frac{\Delta_{\bp}}{E_{\bp}},\qquad
 u_{\bp}^2-v_{\bp}^2=\cos
 2\theta_{\bp}=\frac{\xi_{\bp}}{E_{\bp}}.
\end{equation}
Within the self-consistent BCS approximation the
$\hat\psi$-``particles'' can be considered as noninteracting, and
their thermal-averaged number and entropy are given by the familiar
expressions, respectively,
\begin{gather}
 n_{\bp}\equiv \langle\hat{n}_{\bp\alpha}\rangle
  = \langle \hat\psi^\dagger_{\bp\alpha}\hat\psi_{\bp\alpha}\rangle
  = [\exp(E_{\bp}/\kb T)+1]^{-1},\\
 S(T) = -2\kb\sum_{\bp}\left[n_{\bp}\ln
  n_{\bp}+(1-n_{\bp})\ln (1-n_{\bp})\right].
\label{eq:entropy}
\end{gather}
Similarly, substituting in \Eref{eq:H}
\begin{equation}
 \left(\begin{array}{c}
         \hat{c}^\dagger_{\bp\uparrow}\\
         \hat{c}_{-\bp\downarrow}
       \end{array}\right) =
 \left(\begin{array}{cc}
        u_{\bp} & -v_{\bp}\\
        v_{\bp} &  u_{\bp}
       \end{array}\right)
 \left(\begin{array}{c}
        \hat{\psi}^{\dagger}_{\bp\uparrow}\\
        \hat{\psi}_{-\bp\downarrow}
       \end{array}\right)
\end{equation}
we obtain for the expectation value of the reduced Hamiltonian with
respect to the BCS ground state (\ref{eq:bcs0})
\begin{equation}
 \langle\hat{H}'\rangle =
    2\sum_{\bp} \xi_{\bp}\left[v_{\bp}^2+\left(u_{\bp}^2-v_{\bp}^2\right)
    n_{\bp}\right]
    -\frac{G}{\cal{N}}\left|\sum_{\bp} \chi_{\bp} u_{\bp} v_{\bp} (1-2n_{\bp})\right|^2.
\label{eq:energy}
\end{equation}
The minimization of the free energy
\begin{equation}
 F = \langle\hat{H}'\rangle - TS
\label{eq:free}
\end{equation}
with respect to $\Delta_{\bp}$ leads to the gap equation
(\ref{eq:sdGap}). Then the substitution of the so obtained gap in
\Eref{eq:free} gives the desired form of the minimal free energy. It
should be noted that in order for us to derive the free energy as a
function of the order parameter, $F(\Xi)$, we have to use
\Eref{eq:anzatz} as \emph{ansatz}, thereby considering the order
parameter $\Xi$ as an independent variable for a fixed momentum
dependence of the gap function $\Delta_{\bp}=\Xi\chi_{\bp}$. Thus, the
desired function $F(\Xi,T)$ is obtained from \Eref{eq:free} by
substituting the expression for the averaged energy, \Eref{eq:energy},
and that for the entropy, \Eref{eq:entropy}, which can be also
rewritten in the form
\begin{equation}
 -TS =\sum_{\bp}\left[E_{\bp}\tanh\frac{E_{\bp}}{2\kb T}
      -2\kb T\ln\left(2\cosh\frac{E_{\bp}}{2\kb T}\right)\right].
\end{equation}
The self-consistent BCS approximation gives an analytical dependence
on the order parameter $\Delta_{\bp}=\Xi\,\chi_{\bp}$,
\begin{align}
 \frac{F(\Xi,T)}{\cal{N}} = &
   \, 2 \la
   \xi_{\bp}\left[\frac{1}{2}\left(1-\frac{\xi_{\bp}}{E_{\bp}}\right)
   + \frac{\xi_{\bp}}{E_{\bp}}\,
   \frac{1}{\exp(E_{\bp}/\kb T)+1}
   \right]
   \ra_{\!\!\bp}
 \nn \\
 & \quad - G \left|\la
   \frac{\left|\chi_{\bp}\right|^2}{2 E_{\bp}}\, \tanh\frac{E_{\bp}}{2\kb T}
   \ra_{\!\!\bp}\right|^2\left|\Xi\right|^2
    \nn \\
 & \; \qquad + 2\kb T\la\frac{E_{\bp}}{2\kb T}\, \tanh\frac{E_{\bp}}{2\kb T}
      - \ln\left(2\cosh\frac{E_{\bp}}{2\kb T}\right)\ra_{\!\!\bp}.
\label{eq:FreeEnergy}
\end{align}
Then the Taylor expansion provides the
coefficients in the GL functional per unit cell
\begin{equation}
\frac{F(\Xi,T)}{\cal{N}} \approx a(T) \eta + \frac{1}{2}b(T)\eta^2 +O(\eta^3),
 \qquad \eta \equiv \left|\Xi\right|^2,
\label{eq:GLF}
\end{equation}
which is often a satisfactory approximation even far from the critical
temperature. Let us mention that close to \Tc
\begin{equation}
a(T)\approx a_0\frac{T-\Tc}{\Tc}, \qquad b(T)\approx b\equiv b(T_c) >0.
\label{eq:GLcoeff}
\end{equation}
The order parameter $\Xi,$ being complex in the general case, is
proportional to the GL effective wave function, while $\eta$ is proportional
to the superfluid density.

A straightforward way to obtain the $a_0$ and $b$ coefficients is to
compare the variations with respect to $\eta$ of the
expression (\ref{eq:free}) and the GL expansion
(\ref{eq:GLF}). Introducing the notation $\beta=1/\kb T$ the former
reads
\begin{align}
 \frac{\delta F}{\delta\eta} = &
   \la \frac{\xi^2_{\bp}|\chi_{\bp}|^2}{2E^3_{\bp}}
   \left[1-2n_{\bp} - 2\beta E_{\bp} n_{\bp}(1-n_{\bp})\right]
   \ra_{\!\!\bp}
 + \beta\la n_{\bp}(1-n_{\bp})|\chi_{\bp}|^2\ra_{\!\bp} \nn \\
 & - G\la \frac{|\chi_{\bp}|^2}{2E_{\bp}}(1-2n_{\bp}) \ra_{\!\!\bp}
    \left[\la\frac{|\chi_{\bp}|^2}{2 E_{\bp}}(1-2n_{\bp})
    \ra_{\!\!\bp}\right. \nn \\
 & \qquad\left. -2\eta \la \frac{|\chi_{\bp}|^4}{4E^3_{\bp}}
   \left[1-2n_{\bp} - 2\beta\, E_{\bp} n_{\bp}(1-n_{\bp})\right]
   \ra_{\!\!\bp}\right]. \label{eq:dF}
\end{align}

Furthermore, introducing the functions
\begin{align}
{\cal A}(\eta,T) & = \la
\frac{|\chi_{\bp}|^2}{2E_{\bp}}(1-2n_{\bp})\ra_{\!\!\bp},
 \label{eq:A}\\
{\cal B}(\eta,T) & = \la  \frac{|\chi_{\bp}|^2}{2E^3_{\bp}}\left
         [(1-2n_{\bp})\xi_{\bp}^2
         + 2\beta\, E_{\bp}|\Delta_{\bp}|^2\, n_{\bp}(1-n_{\bp})\right]\ra_{\!\!\bp},
\label{eq:B}
\end{align}
\Eref{eq:dF} can be cast in the compact form
\begin{equation}
 \frac{\delta F}{\delta\eta} = {\cal B}(\eta, T)(1 - G{\cal A}(\eta,
 T) ).
\label{eq:dFshort}
\end{equation}
Notice that the extremum condition of (\ref{eq:dFshort}), $1-G{\cal
A}(\eta,T)=0,$ gives the gap equation (\ref{eq:sdGap}). Now varying
\Eref{eq:GLF} with respect to $\eta$ and comparing with
\Eref{eq:dFshort} we find
\begin{equation}
a_0 = -\Tc\left. \frac{\partial{\cal A}(\eta,T)}{\partial
T}\right|_{\eta=0,\,T=\Tc},\qquad
b = -\left. \frac{\partial{\cal A}(\eta,T)}{\partial
\eta}\right|_{\eta=0,\,T=\Tc},
\label{eq:GLcoeff2}
\end{equation}
where we have used the identity ${\cal B}(0,\Tc)={\cal A}(0,\Tc)=
1/G.$ Thus, \Eref{eq:GLF} with Eqs.~(\ref{eq:GLcoeff}), (\ref{eq:A}),
(\ref{eq:B}), and (\ref{eq:GLcoeff2}) provide the complete set of
equations which determines the GL free energy of an anisotropic-gap
superconductor.

We shall proceed now with working out explicit expressions on the
basis of the general relations (\ref{eq:GLcoeff2}). Taking the
corresponding derivatives of ${\cal A}(\eta,T)$ the $a_0$ and $b$
coefficients read
\begin{equation}
a_0 = \frac{1}{4\kb\Tc}
   \la\frac{|\chi_{\bp}|^2}{\cosh^2\nu_{\bp}}\ra_{\!\!\bp},\qquad
b = \frac{1}{4(\kb\Tc)^3}\la|\chi_{\bp}|^4\,Q(\nu_{\bp})\ra_{\bp},
\label{eq:GLcoeff3}
\end{equation}
where $\nu_{\bp} = \xi_{\bp}/2\kb\Tc$ and (cf. \Rref{Kishore:99})
\begin{equation}
Q(x) = \frac{1}{x^2}\left( \frac{\tanh x}{x} - \frac{1}{\cosh^2
       x}\right) = \frac{\tanh^2 x}{x^2}+\frac{\tanh x - x}{x^3}.
\label{eq:nu_D}
\end{equation}
In calculating the momentum averages (\ref{eq:GLcoeff3}) one has to
take into account that the integrands in the momentum-space integrals
exhibit sharp maxima at the Fermi surface $\eps_{\bp}=E_{\mathrm{F}}.$
Thus, within acceptable accuracy, one can carry out an integration
along the normal to the Fermi surface whereupon $\nu_{\bp}$ may take
on values in the interval $(-\infty,+\infty)$ while the longitudinal
quasimomentum component $p_{\|}$ varies negligibly. Taking into
account the numerical values of the integrals
$\int_{-\infty}^{+\infty} (\cosh x)^{-2} dx = 2,$ and
\begin{equation}
 \int_{-\infty}^{\infty} Q(x) dx =
 8\sum_{n=0}^{\infty}\int_0^{\infty}
 \frac{dx}{[x^2 + \frac{\pi^2}{4}(2n+1)^2]^2} =
 \frac{16}{\pi^2}\sum_{n=0}^{\infty}\frac{1}{(2n+1)^3}=
 \frac{14}{\pi^2}\zeta(3),
\label{eq:auxint}
\end{equation}
where $\zeta(m)$ is the Riemann zeta function,\footnote{This result is
easily obtained by employing the infinite-series representation $
\tanh(x/2)= 4x
\sum_{n=0}^{\infty}\left[\pi^2(2n+1)^2+x^2\right]^{-1},$ and recalling
the relation between the Hurwitz and the Riemann zeta functions,
respectively, $\zeta(m,\frac{1}{2})=\sum_{n=0}^{\infty} 1/(n +
\frac{1}{2})^m = (2^m -1)\zeta(m).$}\ \
we obtain the desired form
\begin{equation}
a_0 = \la \left| \chi_{\bp} \right|^2\ra_{\text{F}}, \qquad
  b = \frac{7 \zeta(3)}{8\pi^2 (\kb\Tc)^2}\la \left|
  \chi_{\bp} \right|^4\ra_{\text{F}}.
\label{eq:final}
\end{equation}
Equation~(\ref{eq:final}) reproduces the results by Pokrovski\u\i\ and
Ryvkin\cite{Pokrovsky:63} and Gor'kov and
Melik-Barkhudarov\cite{Gorkov:64} obtained for the case of a clean
superconductor. The detailed consideration of the influence of
nonmagnetic impurities on the anisotropic order parameter was given by
Pokrovsky and Pokrovsky.\cite{Pokrovsky:96} Very recently
Kogan\cite{Kogan:02} has analyzed the macroscopic anisotropy in
superconductors with anisotropic gap.

For the momentum-space integrals over the Fermi surface which appear
in \Eref{eq:final} we use the notation
\begin{equation}
 \la f(\bp) \ra_{\text{F}} \equiv \sum_b \idotsint
 \delta(\eps_{b,\bp}-E_{\mathrm{F}}) f(\bp) \frac{d\bp}{(2\pi)^D} =
 \sum_b \idotsint\limits_{\eps_{b,\bp} = E_{\mathrm{F}}}f(\bp) \frac{d
 S_{b,\bp}}{v_{b,\bp}(2\pi)^D},
\end{equation}
where $dS_{b,\bp}$ is an infinitesimal surface element of the
(possibly fragmented) Fermi surface sheet of the $b$th energy band,
and $\mathbf{v}_{b,\bp}=\nabla_{\bp}\eps_{b,\bp}$ is the quasiparticle
``velocity'' which according to the present convention has dimension
of energy. Conversion to the true velocity can be performed by
multiplying the (dimensionless) quasimomentum by $\hbar/a,$ where $a$
is the lattice constant. For a conventional electron-phonon pairing
mechanism the isotropic gap, $\chi_{\bp}\equiv 1,$ is often a
reasonable approximation.  Thereby averaging in \Eref{eq:final}
results in the density of states (per spin orientation $\alpha$) at
the Fermi level $\rho_{\mathrm{F}}=\la 1 \ra_{\mathrm{F}}.$ In this
important case the general formulae are in agreement with the
classical result by Gor'kov\cite{Gorkov:59} (in particular,
cf. Eq.~(11) in \Rref{Gorkov:59}).

In thermodynamics of second-order phase transitions the ratio of the
GL coefficients $a_0$ and $b$ determines the jump of the specific
heat\cite{LLV} per unit cell at the critical temperature, so for
superconductors
\begin{equation}
  \frac{1}{\cal N}(C_{\mathrm{s}}- C_{\mathrm{n}})|_{\Tc}
  = \frac{\Delta C}{\cal N} = \frac{1}{\Tc} \frac{a_0^2}{b},
\label{eq:DeltaC}
\end{equation}
where $C_{\mathrm{s}}$ is the specific heat of the superconducting
phase, and
\begin{equation}
 \left.\frac{C_{\mathrm{n}}}{\cal
 N}\right|_{\Tc}=\frac{2}{3}\pi^2\kb^2\rho_{\mathrm{F}}\, \Tc
\end{equation}
is the normal-phase specific heat per unit cell at \Tc; the factor 2
takes into account the spin degeneracy of the normal paramagnetic
phase.  These formulae can be derived directly from the BCS expression
for the specific heat\cite{bcs:57,Zarate:83}
\begin{equation}
 \frac{C(T)}{\cal N}=T\frac{d}{dT}\frac{S(T)}{\cal N}
 =\frac{2}{\kb T^2} \la n_{\bp}(1-n_{\bp})
  \left(E_{\bp}^2-\frac{T}{2}\,
  \frac{d}{dT}|\Delta_{\bp}(T)|^2\right)\ra_{\!\!\bp}.
\label{eq:HeatCapacity}
\end{equation}

Substituting \Eref{eq:final} into (\ref{eq:DeltaC}), and taking into
account that the order parameter $\Xi$ is not affected by momentum
averaging, one finds the general expression for the relative jump of
the specific heat of an anisotropic-gap superconductor
\begin{equation}
 \left.\frac{\Delta C}{C_{\mathrm{n}}}\right|_{\Tc}
 = \frac{12}{7\zeta(3)} \frac{1}{\beta_{\Delta}}, \qquad
   \frac{1}{\beta_{\Delta}}  = \frac{ \la|\Delta_{\bp}|^2\ra_{\text{F}}^2}{%
             \la 1 \ra_{\text{F}} \la |\Delta_{\bp}|^4 \ra_{\text{F}}
} \leqslant 1,\qquad
\frac{12}{7\zeta(3)} = 1.42613\dots,
\label{eq:jump}
\end{equation}
the latter being the universal BCS ratio.\footnote{The numerical value
for the specific heat ``anomaly'' at \Tc\ cited originally by Bardeen,
Cooper and Schrieffer\cite{bcs:57} is slightly different, $\Delta
C/C_{\mathrm{n}}|_{\Tc} = 1.52.$}\ \ Note that $\beta_{\Delta}$ is
similar to the Abrikosov parameter\cite{Abrikosov:88} $\beta_A.$

\section{Applications}
\vspace*{-0.5pt}
\noindent

\subsection{Jump of the specific heat for layered cuprates}

In models for layered cuprates one often postulates the
following functional form of the gap anisotropy
\[
 \Delta_{\bp} \propto \cos (2\phi_{\bp}),
\]
where $\phi_{\bp} = \arctan (p_y/p_x).$ In the case of a parabolic
dispersion, $\eps_{\bp} \propto p^2,$ averaging in \Eref{eq:jump} is
straightforward and yields
\[
  1/\beta_{\Delta} = 2/3.
\]
This value should be considered as a lower bound for
$1/\beta_{\Delta}$ as the Fermi velocity is minimal where the
superconducting gap is maximal. Consistent with that, for a realistic
band dispersion\cite{Mishonov:00} and gap anisotropy\cite{Mishonov:02}
the reduction factor may well reach $1/\beta_{\Delta} = 0.72.$

As pointed out by Abrikosov,\cite{Abrikosov:97} if the Fermi level is
very close to an extended saddle-point singularity the derivation of
the GL equations requires a special consideration.
\pagebreak[3]

\subsection{Jump of the specific heat for MgB$_2$}

Let us now make some comparison between different models applied to
MgB$_2.$
In a very recent e-print Posazhennikova, Dahm and Maki\cite{Dahm:02}
discuss a model for the gap anisotropy in MgB$_2$, a material which
has attracted a lot of attention from condensed-matter physicists in
the past two years.  A central issue in this work\cite{Dahm:02} is to
propose an analytic model for analyzing thermodynamic
behavior. Assuming a spherical Fermi surface, a simple gap anisotropy
function is suggested, $\Delta(\bp)=\Delta_e/\sqrt{1+ A z^2}$, where
$z = \cos\theta$, and $\theta$ is the polar angle.  This model leads
to useful results for the temperature dependence of the upper critical
field $H_{\mathrm{c2}}$ and of the specific heat, which can be fitted
to the experimental data, thereby determining the optimal anisotropy
parameter $A$. Note that $A=(\Delta_e/\Delta_p)^2-1$, with $\Delta_p =
\Delta (z=1)$ and $\Delta_e = \Delta (z=0)$, and the gap ratio is
parameterized as $\Delta_e/\Delta_p=\sqrt{1+A}>0$.

We shall now apply the general results obtained in Sec.~3 to derive a
convenient analytical expression giving the possibility for
determining $\Delta_e/\Delta_p$ from the available data for the jump
of the specific heat.\cite{Wang:01}
\begin{figure}[t]
\centering\includegraphics[width=0.65\textwidth]{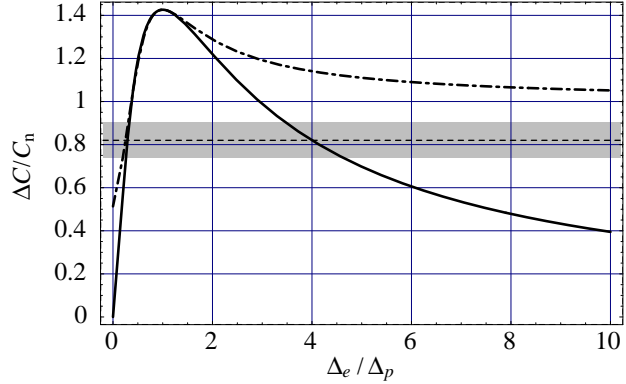}
\caption{\small Jump of the specific heat $\Delta
C/C_{\mathrm{n}}|_{\Tc}$ versus the ``equatorial-to-polar'' gap ratio
$\Delta_e/\Delta_p$. For a given $\Delta C/C_{\mathrm{n}}|_{\Tc}$
value we have oblate $\Delta_e/\Delta_p>1$ and prolate
$\Delta_e/\Delta_p<1$ solutions. The solid
curve\protect\cite{MPI:02_b} represents \protect\Eref{eq:oblate}, and
is applicable to the model by Posazhennikova, Dahm and
Maki.\protect\cite{Dahm:02} The dash-dotted line is our analytical
solution,\protect\cite{MPI:02} \Eref{eq:1band}, applicable for the
model by Haas and Maki.\protect\cite{Haas:01} The dashed line is the
jump ratio $\Delta C/C_{\mathrm{n}}|_{\Tc}=0.82\pm10\%$ measured by
Wang, Plackowski, and Junod,\protect\cite{Wang:01} with the shaded
area showing the experimental error bar.
\label{fig:1}
}
\end{figure}
Following the weak-coupling BCS approach,\cite{Dahm:02,Haas:01} from
\Eref{eq:jump} we find\cite{MPI:02_b} for $A>0,$ and $-1<A<0$, respectively,
\newcommand{\D}{\displaystyle}
\begin{equation}
\left.\frac{\Delta C(A)}{C_{\mathrm{n}}}\right|_{\Tc} =
\begin{cases}
  \frac{\D 12}{\D 7\zeta(3)}
      \frac{\D 2(1 + w^2)\left(\arctan w\right)^2}
           {\D w^2 + w(1 + w^2)\arctan w}, &
 w = \sqrt{A}=\sqrt{\left(\frac{\Delta_e}{\Delta_p}\right)^2-1},
\label{eq:oblate}\\[0.4cm]
 \frac{\D 12}{\D 7\zeta(3)}
      \frac{\D 2(1-y^2)\left(\tanh^{-1} y\right)^2}
           {\D y^2 + y(1-y^2)\tanh^{-1} y}, &
 y = i w = \sqrt{-A}=\sqrt{1-\left(\frac{\Delta_e}{\Delta_p}\right)^2}.
\end{cases}
\end{equation}
For a given specific-heat jump, this expression leads to \emph{two}
solutions (oblate, $\Delta_e/\Delta_p>1,$ and prolate,
$\Delta_e/\Delta_p<1$). The relevant example is shown in \Fref{fig:1}.
In \Rref{Dahm:02} values of $\Delta C(A)/C_{\mathrm{n}}|_{\Tc}$
consistent with \Eref{eq:oblate} have been tabulated. The analysis of
the experimental data\cite{Xu:01,Angst:02} for the angular dependence
of $H_{\mathrm{c2}}$ unambiguously demonstrates that one has to
analyze only the ``oblate'' case. The experimental value $\Delta
C/C_{\mathrm{n}}|_{\Tc}=0.82 \pm 10\%,$ reported in \Rref{Wang:01},
gives $A\approx 16$ and $\Delta_e/\Delta_p\approx w \approx 4.0 \pm
10\%$.
For this significant anisotropy, the ``distribution'' of Cooper pairs
$
\langle\left|\Delta_{\bp}\right|^2\rangle \propto 1/[p_z^2+(p_{\mathrm{F}}/w)^2]
$
has a maximum at $p_z=0,$ and $p_{\mathrm{F}}$ is the Fermi
momentum.  This general qualitative conclusion is in agreement with
the hints from band calculations that the maximal order parameter is
concentrated in a nearly two-dimensional electron band, but all bands
$\eps_{b,\bp}$ contribute to $C_{\mathrm{n}}/{\cal N}.$

In another paper Haas and Maki\cite{Haas:01} considered the model gap
anisotropy $\Delta\propto 1 + a z^2$ for which similar calculation
gives\cite{MPI:02}
\begin{align}
\left.\frac{\Delta C}{C_{\mathrm{n}}}\right|_{\Tc}
  & = \frac{12}{7\zeta(3)}
      \frac{1+ 4 a /3 + 38 a^2/45 + 4 a^3/15 + a^4 /25}
      {1+ 4 a/3 + 6 a^2/5+ 4 a^3/7 + a^4/9},\nonumber \\
a & =\frac{1}{\Delta_e/\Delta_p}-1,\qquad
     \Delta_e/\Delta_p=\frac{1}{1+a}.
\label{eq:1band}
\end{align}
This model, however, cannot explain the significant reduction of
$\Delta C/C_{\mathrm{n}}|_{\Tc}$ observed experimentally, as can be
seen in \Fref{fig:1}.

\subsubsection{$\Delta C/C_{\mathrm{n}}|_{\Tc}$ within a two-band model}

\begin{figure}[t]
\centering\includegraphics[width=0.65\textwidth]{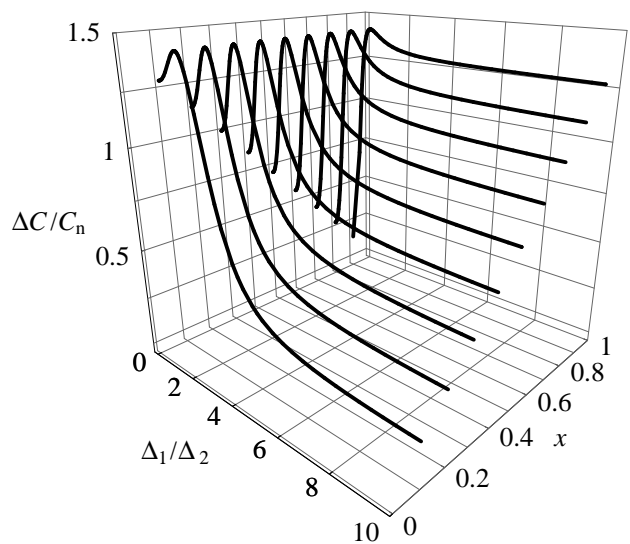}
\caption{\small Jump of the specific heat $\Delta C/C_{\mathrm{n}}$ at
$T_c$ versus the gap ratio $\Delta_1/\Delta_2$ for a set of different
$x$ values.  The latter parameter is the relative contribution of the
``first'' band to the density of states within the two-band model for
MgB$_2$.}
\label{fig:2}
\end{figure}

For the Moskalenko\cite{Moskalenko:59}-two-band model, advocated for
the first time for MgB$_2$ in \Rref{Shulga:01}, \Eref{eq:jump} gives
(to within a typographical correction) the result by
Moskalenko,\cite{Moskalenko:59}
\begin{align}
 \left.\frac{\Delta C}{C_{\mathrm{n}}}\right|_{\Tc} &= \frac{12}{7\zeta(3)}\,
 \frac{(|\Delta_1|^2 \rho_1+ |\Delta_2|^2 \rho_2)^2}
 {(\rho_1+\rho_2)(|\Delta_1|^4 \rho_1+ |\Delta_2|^4 \rho_2)}\nn \\
 & =1.426\,\frac{[\delta^2 x + (1 - x)]^2}{\delta^4 x+(1-x)}<1.43,
\label{eq:2band}
\end{align}
where
$$
 \delta = \frac{\Delta_1}{\Delta_2},\qquad
 x = \frac{\rho_1}{\rho_1+\rho_2},
$$
and $\rho_1$ and $\rho_2$ are the densities of states for the two
bands at the Fermi level. In \Fref{fig:2} we present the specific heat
jump as a function of $\delta$ for various values of $x.$ Taking for
an illustration $x=0.515$ and $\Delta C/C_{\mathrm{n}}|_{\Tc}=0.82,$
\Eref{eq:2band} gives $\Delta_1/\Delta_2\approx 4.0$ in agreement with
the maximal-to-minimal-gap ratio $\Delta_e/\Delta_p \approx 4.0$
obtained using \Eref{eq:oblate}.

We should stress here that the two-band model also describes a kind of
``aniso\-tro\-py'' in the sense that a ``\emph{non-constant}'' order
parameter, having \emph{different constant values} within the
different bands comprising the Fermi surface, leads to modified GL
parameters and reduces the jump of the specific heat.  For a survey on
a set of parameters see Table~I in \Rref{Bouquet:01}. Certainly, from
the jump of the specific heat alone one cannot judge about the
validity of any model, so subtleties, e.g., related to strong coupling
effects and other anisotropies, can be hidden in the spread of the
parameters in the table mentioned.

As we have aimed here on methodological aspects as well, it is beyond
the scope of this work to analyze in detail different experimental
data and the theoretical fits to them. Nonetheless we shall mention
that the Moskalenko-two-band model\cite{Moskalenko:59} with isotropic
gaps agrees better\cite{Bouquet:01} with the experimental data for
MgB$_2$ compared to other proposed models, but it would be premature
to make any final judgment. Thus the synthesis of crystals with really
sharp transition may well be rendered necessary.

\subsection{Effect of finite Debye frequency $\omega_{\mathrm{D}}$ for phonon
superconductors}

Up to now we have dealt with energy scales corresponding to very high
phonon frequencies or exchange interaction.  For phonon
superconductors, however, we have to take into account only a narrow
``layer'' of pairing electrons with energies $|\xi_{\bp}|<
\hbar\omega_{\mathrm{D}}$. In this case $\nu_{\bp} \in
(-\hbar\omega_{\mathrm{D}}/2\kb\Tc,+\hbar\omega_{\mathrm{D}}/2\kb\Tc),$
and according to \Eref{eq:final} we have
\begin{align}
a_0 =& \la \left| \chi_{\bp} \right|^2\ra_{\text{F}}
 \tanh\frac{\hbar\omega_{\mathrm{D}}}{2\kb\Tc},\nn \\
  b =& \frac{1}{8(\kb\Tc)^2}
  \la \left|
  \chi_{\bp}
 \right|^4\ra_{\text{F}}\int_0^{\hbar\omega_{\mathrm{D}}/2\kb\Tc}Q(x)dx.
\label{eq:fin}
\end{align}
For the relative jump of the specific heat, using \Eref{eq:DeltaC} and
\Eref{eq:jump}, we obtain
\begin{equation}
 \left.\frac{\Delta C}{ C_{\mathrm{n}}}\right|_{\Tc}
 = \frac{12}{\pi^2}\,\frac{1}{\beta_{\Delta}}\,
           \frac{\tanh^2(\hbar\omega_{\mathrm{D}}/2\kb\Tc)}
                {\int_0^{\hbar\omega_{\mathrm{D}}/2\kb\Tc}Q(x)d x}.
\label{eq:jumpphonon}
\end{equation}
Kishore and Lamba\cite{Kishore:99} have recently shown that the
behavior of $\Delta C/\Tc$ calculated within the BCS model with finite
$\omega_{\mathrm{D}}$ is very similar to the results by
Marsiglio~\etal\cite{Marsiglio:87} and Carbotte\cite{Carbotte:90}
obtained from the Eliashberg theory; see also \Rref{Marsiglio:88} and
Refs.~\citen{Muhlschlegel:59,Clem:66}.  As regards the recent
first-principles linear-response calculation\cite{Golubov:02} for
MgB$_2$ the agreement between the theory and experiment is comparable
to the quality of the fit\cite{Bouquet:01} to experimental data
employing the original week-coupling two-band
model.\cite{Moskalenko:59} Precise analysis, however, has
manifested\cite{Kortus:02} a significant temperature dependence of
$\Delta_1/\Delta_2$, which we interpret as an evidence for strong
coupling effects. Further support can be possibly gained from the
direct comparison between the calculated relative jump and the week
coupling formula \Eref{eq:2band}.

\section{Summary}

Analyzing the Ginzburg-Landau coefficients and the jump of the
specific heat at \Tc\ we were able to establish that a number of
special cases discussed in the recent literature on superconducting
cuprates, MgB$_2$ and borocarbides can be easily derived as a
consequence of the classical treatment of thermodynamics of
anisotropic superconductors carried out in the dawn of the BCS era. It
is somewhat strange that all those early papers are not being used in
the current literature and a number of results are rediscovered or
treated by a numerical brute force. We have cast, where necessary with
the proper correction(s), those classical results into a form to be
used for fitting the experimental data. The relative jump of the
specific heat is one of the most appropriate quantities which can
provide important information for the properties of
superconductors. Thus, it would be interesting to see an analogous
approach for unconventional superconductors\cite{Sauls:94} as well.

\nonumsection{Acknowledgments}
\noindent
The authors gratefully acknowledge the stimulating discussions with
Joseph Indekeu, his hospitality and interest in this study as well as
his critical reading of the manuscript and the great number of
valuable improvements suggested. One of the authors (T.~M.) is
thankful to A.~A.~Abrikosov, T.~Dahm, S.-L.~Drechsler,
O.~Entin-Wohlman, L.~P.~Gor'kov, A.~Junod, R.~Kishore, J.~Kortus,
K.~Maki, V.~A.~Moskalenko, M.~E.~Palistrant, and V.~L.~Pokrovsky for
the clarifying correspondence related to their papers. This work was
supported by the Belgian DWTC, the Flemish Government Programme
VIS/97/01, the IUAP and the GOA.

%
\nonumsection{References}

\end{document}